\documentclass[11pt,twoside]{article}
\usepackage{asp2014}
\newbox\grsign \setbox\grsign=\hbox{$>$} \newdimen\grdimen
\grdimen=\ht\grsign
\newbox\simlessbox \newbox\simgreatbox \newbox\simpropbox
\setbox\simgreatbox=\hbox{\raise.5ex\hbox{$>$}\llap
     {\lower.5ex\hbox{$\sim$}}}\ht1=\grdimen\dp1=0pt
\setbox\simlessbox=\hbox{\raise.5ex\hbox{$<$}\llap
     {\lower.5ex\hbox{$\sim$}}}\ht2=\grdimen\dp2=0pt
\setbox\simpropbox=\hbox{\raise.5ex\hbox{$\propto$}\llap
     {\lower.5ex\hbox{$\sim$}}}\ht2=\grdimen\dp2=0pt

\aspSuppressVolSlug
\resetcounters

\begin{document}

\title{Refining our knowledge of the white dwarf mass-radius relation with HST observations of Sirius-type binaries}

\author{M.A. Barstow,$^{1}$, H. E. Bond, $^{2}$, M.R. Burleigh, $^{1}$, S.L. Casewell,$^1$, J. Farihi,$^{3}$, J.B. Holberg,$^{4}$ and I. Hubeny,$^{5}$}
\affil{$^1$Department of Physics and Astronomy, University of Leicester, University Road, Leicester, LE1 7RH, UK; \email{mab@le.ac.uk}}
\affil{$^2$Department of Astronomy and Astrophysics, Pennsylvania State University, University Park, PA16802, USA}
\affil{$^3$Department of Physics and Astronomy, University College London, Gower street, London, WC1E 6BT, UK}
\affil{$^4$Lunar and Planetary Laboratory, University of Arizona, Tuscon, AZ 85721, USA}
\affil{$^5$Steward Observatory, University of Arizona, 933 N. Cherry Tree Avenue, Tuscon, AZ 85721, USA}
\paperauthor{M.A. Barstow}{mab@le.ac.uk}{}{University of Leicester}{Department of Physics and Astronomy}{Leicester}{Leics}{LE1 7RH}{UK}
\paperauthor{H.E. Bond}{heb11@psu.edu}{}{Pennsylvania State University}{Department of Astronomy and Astrophysics}{}{}{PA16802}{USA}
\paperauthor{M.R. Burleigh}{mbu1@le.ac.uk}{}{University of Leicester}{Department of Physics and Astronomy}{Leicester}{Leics}{LE1 7RH}{UK}
\paperauthor{S.L. Casewell}{slc25@le.ac.uk}{}{University of Leicester}{Department of Physics and Astronomy}{Leicester}{Leics}{LE1 7RH}{UK}
\paperauthor{J. Farihi}{jfarihi@star.ucl.ac.uk}{}{University College London}{Department of Physics and Astronomy}{London}{}{WC1E 6BT}{UK}
\paperauthor{J.B. Holberg}{holberg@vega.lpl.arizona.edu}{}{University of Arizona}{Lunar Planetary Laboratory}{Tuscon}{}{AZ85721}{USA}
\paperauthor{I. Hubeny}{hubeny@as.arizona.edu}{}{University of Arizona}{Steward Observatory}{Tuscon}{}{AZ85721}{USA}

\begin{abstract}
The presence of a white dwarf in a resolved binary system, such as Sirius, provides an opportunity to combine dynamical information about the masses, from astrometry and spectroscopy, with a gravitational red-shift measurement and spectrophotometry of the white dwarf atmosphere to provide a test of theoretical mass-radius relations of unprecedented accuracy. We demonstrated this with the first Balmer line spectrum of Sirius B to be obtained free of contamination from the primary, with STIS on HST. However, we also found an unexplained discrepancy between the spectroscopic and gravitational red-shift mass determinations. With the recovery of STIS, we have been able to revisit our observations of Sirius B with an improved observation strategy designed to reduce systematic errors on the gravitational red-shift measurement. We provide a preliminary report on the refined precision of the Sirius B mass-radius measurements and the extension of this technique to a larger sample of white dwarfs in resolved binaries. Together these data can provide accurate mass and radius determinations capable of testing the theoretical mass-radius relation and distinguishing between possible structural models.
\end{abstract}

\section{Introduction}
Evolutionary model calculations (e.g. \citealt{wood95,fontaine01,bergeron01,althaus98,althaus97}) have established that the theoretical radii for DA white dwarfs are systematically dependent on stellar temperature, H layer mass and core composition. However, basic tests based on analyses of a combination of Balmer-line measurements and Hipparcos parallaxes \citep{vauclair97,barstow97,provencal98} are restricted to the narrow range of masses of the isolated white dwarfs (near 0.6 M$_{\odot}$) and to the cooler end of the DA white dwarf T$_{\rm eff}$ range ($<$20,000K), where the M-R relation is less sensitive to temperature/layer  mass and core effects. The hotter stars in the Hipparcos sample (e.g. Feige 24, G191-B2B), which lie at the greatest distances, have parallax uncertainties too large to provide a really thorough examination of the effects of differences in the core and envelope compositions. However, the presence of a bright companion, where the white dwarf resides in a binary system, yields a more accurate parallax. Hence, Sirius-like binary systems present a good opportunity of studying the M-R relation for a range of white dwarf temperatures and masses.

In HST observing cycle 12 we obtained a high-S/N optical spectrum of the Balmer lines of Sirius B with HST/STIS, to provide precise measurements (the first with modern electronic detectors) of T$_{\rm eff}$, log g and gravitational redshift and test the theoretical white dwarf mass-radius (M-R) relation \citep{barstow05}. In cycle 19 we were granted more time to extend this work. We observed Sirius B again, to better understand potential systematic errors in the measurement of its gravitational redshift and also studied five other white dwarfs in resolved Sirius-like systems. We have obtained the first precise measurements of T$_{\rm eff}$, log g and radial velocity for these WDs and combined them with the distances (from Hipparcos) and the primary radial velocities to measure the WD gravitational redshift. As a result we have obtained the stellar mass and radius, to test the M-R relation at high temperatures across a range of white dwarf masses.

\section{Observations}

We have followed the basic instrumental setup used in \citet{barstow05}, and taken observations using the G430L and G750M gratings to cover the full Balmer line series. With its higher resolution, the G750M grating also provides an accurate radial velocity measurement (using the H$\alpha $ lines) and, as a result, gravitational redshift. In the original programme a 52x0.2" slit was used but this appeared to generate systematic uncertainties in both knowledge of the flux (possible light losses from the slit) and velocity measurement (uncertainty in positioning of the stellar image in the slit). Therefore, we used a wider slit for the G430L observation to ensure an accurate flux measurement and recorded a pair of observations in the G750M grating using a wide slit, 52x2.0" for the photometry and normalisation of the data, and a narrow slit, 52x0.05" to obtain a more accurate radial velocity. 

We observed a total of five white dwarfs in Sirius-like systems, SiriusB, HD2133B, 14 Aur BCB, HR1358B and HD217411B. HD217411 was found to be a hierarchical triple system, with the white dwarf lying in an unresolved binary companion to the main sequence star. Therefore, an uncontaminated Balmer line spectrum could not be obtained for the white dwarf and no M-R measurement made. Our findings are published in  \citep{holberg14}. 

All of the spectra were reduced with the \textsc{calstis} pipeline in \textsc{iraf} and analysed as described in \citep{barstow05}. Briefly, we used the spectral fitting code \textsc{xspec} in conjunction with a grid of pure hydrogen synthetic spectra, generated using \textsc{tlusty} and \textsc{synspec}, to obtain measurements of T${\rm eff}$ and log $g$ for each white dwarf. The stellar radial velocity was obtained from the redshift of the model spectrum compared to the rest wavelength of H $\alpha $.

\section{Results}
There are three different approaches that can be taken to measure the mass and radius for a white dwarf, depending on the data available.
The first involves simply taking the T$_{\rm eff}$ and log $g$ as measured from the Balmer line spectrum obtained in the G430L grating and using them as an input to a theoretical mass-radius relation, such as those provided by \citet{fontaine01} to calculate the mass and radius of the white dwarf. 

The second measurement involves calculating the mass directly from the log $g$ measurement as,
\begin{equation}
g=\frac{GM}{R^2}
\end{equation}
where G is the universal gravitational constant, and the radius can be calculated from the normalisation to the model (as STIS spectra are accurately flux calibrated), and the distance to the system.
This measurement is dominated by errors on the system parallax which is known to high accuracy for Sirius B, but much less well for the other systems in our programme. The ESA Gaia mission should produce new measurements before 2017 which will improve the accuracy of our results using this method.

The third method involves determining the gravitational redshift. This is done by measuring the radial velocity of the white dwarf from the H$\alpha $ line and subtracting the $K$ and $\gamma$ velocities of the system.
The gravitational redshift can then be used to determine the mass of the system as, 
\begin{equation}
V_{gr} = \frac{0.636 M}{R}.
\end{equation}
The radius is again calculated from the normalisation of the model to the data. As we have two sets of data - the Balmer lines and the H${\alpha}$ line, we have two measurements of the radius.
\begin{figure}

\centering
\includegraphics[scale=0.3, angle=270]{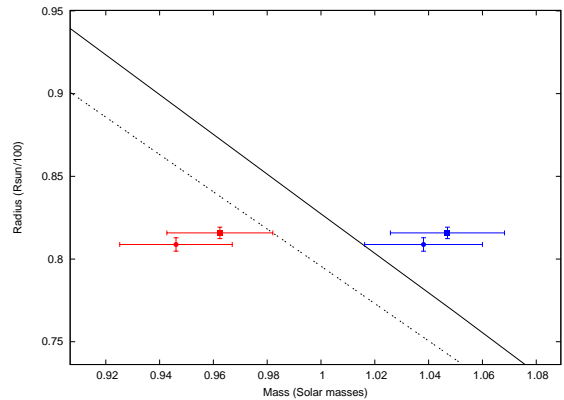}
\caption{The points plotted on the left are the masses derived from the log $g$, while the points on the right are the masses derived from the gravitational redshift. Both sets of radii are derived from the normalisation of the spectra. The squares are the normalisation from the G750M grating and the circles from the G430L normalisation. The solid black line is the  CO core thick H layer and the dotted line is the thin H layer at 25~000 K.}
\label{mr}

\end{figure}

Figure \ref{mr} shows the mass and radius derived from the gravitational redshift (right hand error bars) and from the log $g$ (left error bars), both combined with the photometric normalisation. The square symbols are where the normalisation is from the G750M grating and the circles are from the G430L grating. The diagonal black lines are theoretical mass-radius relations for CO core models at 25~000 K with thick (solid) and thin (dashed) H layers. It can be seen, as also found by \citep{barstow05}, that these two methods of deriving the masses do not agree with each other. Therefore, the refined observational approach designed to remove systematic errors has not resolved the issue and that some problem remains.

\begin{figure}

\centering
\includegraphics[scale=0.25, angle=90]{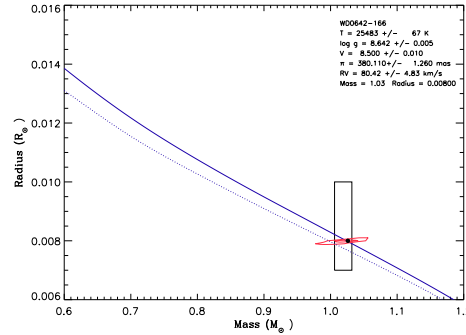}
\caption{The best-fit mass and radius (black dot with 1 and 2 $\sigma $ error ellipses) compared with the astrometric constraints (box) and theoretical M-R relationships for a thick H layer (solid curve) and thin H later (dashed curve). The input parameters values are in the top right corner of the figure.}
\label{fig1}

\end{figure}
Nevertheless, while there is still an analysis issue to be resolved, it is possible to combine all the methods of determining mass and radius and use a chi-squared minimisation to find the best estimate of these parameters. This is illustrated in figure \ref{fig1}, which shows the best fit (black dot) and the 1 and 2 $\sigma $ error ellipses. The diagonal curves are as described in figure \ref{mr} and the box defines the astrometric limits (Bond, private communication). The input parameter values are in the top right corner of the figure. This illustrates that the precision of our measurements allows us to distinguish between theoretical M-R relations assuming different H layer masses.

 \begin{figure}
 \centering
\includegraphics[scale=0.3, angle=270]{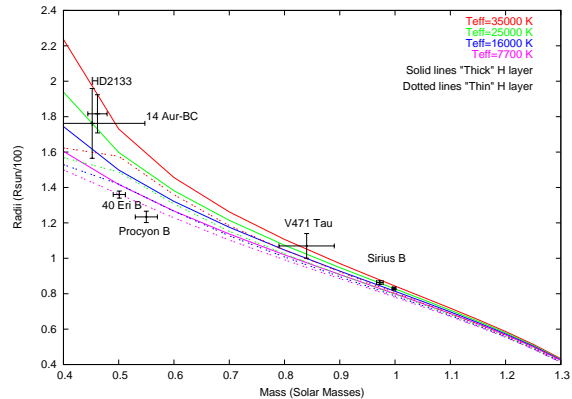}
 \caption{Masses and Radii derived from our STIS data. The Fontaine models \citep{fontaine01} for CO cores and thick and thin H layers are shown for a range of T$_{\rm eff}$. The result for Sirius B in \citet{barstow05} is also shown to the left of our new point which has smaller errors.}
 \label{fig2}
\end{figure}

Figure \ref{fig2} shows the mass-radius relation for 14 Aur C, Sirius B and HD2133 measured as part of this programme. The measurements for Sirius B are derived from  T$_{\rm eff}$ and log $g$ and combined with a mass-radius relation from \citet{fontaine01}.  The mass has increased from our previous work \citep{barstow05}, but it can be seen that our errors have decreased significantly. The measurements for the other stars are from the log $g$ derived mass, and the errors on the parallax dominate.
At this stage, we do not have all the information for the other stars that we have for Sirius B. However, we can illustrate the potential outcome of the programme using a theoretical mass-radius relation from \citep{fontaine01} to anchor the mass and radius measurements. Figure 3 shows the results for 14 Aur C, Sirius B and HD2133, compared to curves for a range of temperatures and thick and thin H layer masses. For HD2133 and 14Aur-BC, the error bars are dominated by the current uncertainties in the system parallax. Nevertheless, the HD2133 result is capable of distinguishing between the models, helped by the increased dynamic range of the models at lower masses. Two results are plotted for Sirius B, from \citep{barstow05} (left) and this work (right). The mass estimate has increased from our previous work, but it can be seen that our errors have also decreased significantly, allowing clear constraints to be placed on the models.

\section{Conclusions}
We have presented preliminary results from our HST/STIS observations of a sample of hot DA white dwarfs residing in resolved Sirius-type binary systems, with the aim of testing the theoretical mass-radius relation under a range of conditions. We have refined our observational technique, but this has not eliminated disagreements in the measurements of M and R from different approaches. It seems most likely that the issue is in the stellar atmosphere models used to fit the Balmer lines rather than the observations themselves. We have demonstrate the ultimate accuracy of our approach but further work is needed to refine some observations, such as the system parallaxes, and obtain more dynamical data for these systems.
\acknowledgements  MAB acknowledges support from the GAIA post-launch support programme of the UK Space Agency. SLC acknowledges support from the College of Science and Engineering at the University of Leicester. 
\bibliography{Mbarstow1}  


\end{document}